\documentclass[twocolumn,english,superscriptaddress,citeautoscript,showpacs,preprintnumbers,amsmath,amssymb,prd,floatfix,nofootinbib]{revtex4-1}
\pdfoutput=1
\usepackage{amsmath, amsthm, amssymb, graphicx,bm,amstext,subfigure, mathdots}
\usepackage{hyperref}
\usepackage{color}

\begin{document}
\title{Reformulation of the Cosmological Constant Problem}
\author{Qingdi Wang}
\affiliation{Department of Physics and Astronomy, 
The University of British Columbia,
Vancouver, BC V6T 1Z1, Canada}
\begin{abstract}
The standard formulation of the cosmological constant problem is based on one critical assumption---the spacetime is homogeneous and isotropic, which is true only on cosmological scales. However, this problem is caused by extremely small scale (Planck scale) quantum fluctuations and at that scale, the spacetime is highly inhomogeneous and anisotropic. The homogeneous Friedmann-Lema\^{i}tre-Robertson-Walker metric used in the standard formulation is inadequate to describe such small scale dynamics of the spacetime. In this Letter, we reformulate the cosmological constant problem by using a general inhomogeneous metric. The fine-tuning problem does not arise in the reformulation since the large gravitational effect of the quantum vacuum is hidden by small scale spacetime fluctuations. The stress energy tensor fluctuations of the quantum fields vacuum could serve as ``dark energy" to drive the accelerating expansion of the Universe through a weak parametric resonance effect.
\end{abstract}
\maketitle

\section{Introduction}\label{problem}
The cosmological constant problem is a fundamental problem in modern physics. Since the theoretical prediction is different from the experimental observation by some 120 orders of magnitude, this problem has been called ``the worst theoretical prediction in the history of physics." Its importance has been emphasized by various authors from different aspects. For example, it has been described as a veritable crisis \cite{RevModPhys.61.1}, an unexplained puzzle \cite{Kolb:EU90}, the most striking problem in contemporary physics \cite{Dolgov:1997za}, and even the mother of all physics problems, the worst prediction ever \cite{citeulike:430764}. It is widely regarded as one of the major obstacles to further progress in fundamental physics \cite{witten}.

This problem arises at the intersection between quantum mechanics and general relativity. It originates from two fundamental principles: (i) the uncertainty principle of quantum mechanics predicts that the quantum fields vacuum possesses a huge amount of zero-point energy; (ii) the equivalence principle of general relativity requires that every form of energy must gravitate in the same way. 

Therefore, the huge energy of quantum vacuum must gravitate to produce a large gravitational effect. In principle, one needs a quantum theory of gravity to study this supposed large gravitational effect. Unfortunately, no satisfactory quantum theory of gravity exists yet.

In order to proceed, the standard formulation of the cosmological constant problem still treats gravity as classical. It assumes the semiclassical Einstein equations,
\begin{equation}\label{Einstein equation usual form}
G_{ab}+\lambda_B g_{ab}=8\pi G \langle T_{ab}\rangle,
\end{equation}
where $\lambda_B$ is the bare cosmological constant and the source of gravity is the expectation value of the quantum vacuum stress energy tensor. Lorentz invariance requires that in the vacuum $\langle T_{ab}\rangle$ takes the form
\begin{equation}\label{vacuum equation of state}
\langle T_{ab}\rangle=-\langle\rho\rangle g_{ab},
\end{equation}
where $\rho$ is the vacuum energy density. The conservation of the stress energy tensor 
\begin{equation}
\nabla^aT_{ab}=0
\end{equation}
requires that $\langle\rho\rangle$ has to be a constant. Then the gravitational effect of the vacuum would be equivalent to a cosmological constant that the Einstein equations \eqref{Einstein equation usual form} can be written as
\begin{equation}\label{vacuum einstein equation}
G_{ab}+\lambda_{\text{eff}} g_{ab}=0,
\end{equation}
where the effective cosmological constant $\lambda_{\text{eff}}$ is defined by
\begin{equation}\label{contributions to effective cc new}
\lambda_{\text{eff}}=\lambda_B+8\pi G\langle\rho\rangle.
\end{equation}

The standard effective field theory arguments predict that, in general, $\langle\rho\rangle$ takes the form
\begin{equation}
\langle\rho\rangle\sim\Lambda^4,
\end{equation}
if one trusts the theory up to a certain high energy cutoff $\Lambda$. This result could have been guessed by dimensional analysis, and the numerical constants that have been neglected will depend on the precise knowledge of the fundamental fields under consideration \cite{Carroll:2000fy}. The cutoff $\Lambda$ is usually taken to be at the Planck energy, i.e., $\Lambda=1$ in Planck units. Then the magnitude of the expectation value of the vacuum energy density
\begin{equation}\label{theorectical value}
\langle\rho\rangle\sim 1,
\end{equation}
where Planck units have been used for convenience.

Note that, in this standard formulation of the cosmological constant problem, the quantum fields vacuum is still treated as classical. It is actually modeled as a perfect fluid, which satisfies the vacuum equation of state \eqref{vacuum equation of state}.

$\langle\rho\rangle$ is generated by very small scale (Planck scale) quantum zero-point fluctuations. It contributes to the effective cosmological constant $\lambda_{\mathrm{eff}}$ through \eqref{contributions to effective cc new}. Thus, $\lambda_{\mathrm{eff}}$ is essentially a quantity meaningful only at small scales; it is not directly observable at macroscopic scales. 

However, in the standard formulation of the cosmological constant problem, $\lambda_{\mathrm{eff}}$ is directly observable at cosmological scale. That is because the standard formulation makes a crucial assumption---the spacetime is homogeneous and isotropic. Then one can use the standard Friedmann-Lema\^{i}tre-Robertson-Walker (FLRW) metric of cosmology to describe the spacetime dynamics given by the Einstein equations \eqref{vacuum einstein equation},
\begin{equation}\label{FLRW}
ds^2=-dt^2+a^2(t)\left(dx^2+dy^2+dz^2\right).
\end{equation}
The solution to \eqref{vacuum einstein equation} under this metric gives an accelerating universe
\begin{equation}
a(t)=a(0)e^{Ht},
\end{equation}
where
\begin{equation}\label{relation}
H=\frac{\dot{a}}{a}=\pm \sqrt{\frac{\lambda_{\mathrm{eff}}}{3}}
\end{equation}
describes the rate of acceleration of the universe. It is either expanding (``$+$" sign) or contracting (``$-$" sign) depending on the initial conditions. In this way, $\lambda_{\mathrm{eff}}$ is related to the macroscopic observable $H$ through \eqref{relation}.

The current Hubble expansion rate $H_0$ provides an upper bound for $\lambda_{\mathrm{eff}}$,
\begin{equation}\label{old prediction}
\lambda_{\mathrm{eff}}=3H^2\leq 3H_0^2\sim 10^{-122},
\end{equation}
where Planck units have been used for convenience.

Therefore, $\lambda_{\mathrm{eff}}$ is different from $\langle\rho\rangle$ by $122$ orders of magnitude, and one has to fine-tune the bare cosmological constant $\lambda_B$ to a precision of 122 decimal places to cancel $\langle\rho\rangle$ to match the observation. This problem of extreme fine-tuning is the so-called cosmological constant problem \cite{RevModPhys.61.1}. It has been described as ``the largest discrepancy between theory and experiment in all of science" \cite{doi:10.1119/1.17850}.

\section{The fluctuating spacetime} 
This large discrepancy depends on the theoretical prediction \eqref{relation} and \eqref{old prediction}, which is based on the homogeneity and isotropy assumption. This assumption is reasonable at cosmological scale, but is not true at small scales (Planck scale). Note that this problem is generated by small scale quantum fluctuations; there is no reason to expect that the cosmological FLRW metric \eqref{FLRW} is still applicable at such small scales. In fact, at small scales the spacetime metric is wildly fluctuating. It is highly inhomogeneous and anisotropic for the following reasons.

First, the vacuum energy density $\rho$ is not a constant because the vacuum is not an eigenstate of the energy density operator $T_{00}$, although it is an eigenstate of the Hamiltonian $\mathcal{H}=\int dx^3 T_{00}$, which is an integral of $T_{00}$ over the whole space. So the average energy density over a relatively large length scale ($\gg 1/\Lambda$) is nearly constant, but, at small length scales ($\sim 1/\Lambda$), $\rho$ cannot be a constant; it is always fluctuating. In fact, the magnitude of the fluctuation is as large as its expectation value \cite{PhysRevD.95.103504}
\begin{equation}
\Delta\rho\sim\langle \rho\rangle.
\end{equation}
More detailed analysis shows that these fluctuations are highly inhomogeneous at small scales \cite{PhysRevD.95.103504}. In general, since the vacuum state is not an eigenstate of the stress energy tensor operator $T_{ab}$, the whole stress energy tensor would be violently fluctuating and highly inhomogeneous. 

Then the resulting spacetime sourced by such wildly fluctuating and highly inhomogeneous vacuum is by no means homogeneous. Moreover, besides these ``passive" fluctuations driven by the fluctuations of the matter field vacuum stress energy tensor, the spacetime also experiences ``active" fluctuations due to the quantum nature of gravity itself \cite{Ford:2000vm, Hu:2008rga}. These fluctuations were already anticipated by John Wheeler \cite{PhysRev.97.511, Wheeler:1957mu, misner1973gravitation} in 1955 that over sufficiently small distances and sufficiently small brief intervals of time, the ``very geometry of spacetime fluctuates." The spacetime would have a foamy, jittery nature and would consist of many small, ever-changing, regions. This picture of highly inhomogeneous fluctuating spacetime is called ``spacetime foam".

Therefore, the standard formulation of the cosmological constant problem that is based on the homogeneous and isotropic FLRW metric is questionable at best. A reformulation of this problem by using a general inhomogeneous metric is needed. 

\section{The reformulation}
We start from the most general metric that is used in the initial value formulation of general relativity:
\begin{equation}\label{new metric}
ds^2=-N^2dt^2+h_{ab}(dx^a+N^adt)(dx^b+N^bdt),
\end{equation}
where the lapse function $N$, the shift vector $N^a$ and the spatial metric $h_{ab}$ depend on both the temporal coordinate $t$ and the spatial coordinate $\mathbf{x}=(x^1, x^2, x^3)$. This metric gives a $3+1$ decomposition of the spacetime by slices $\Sigma_t$ defined by $t=\mathrm{constants}$. It describes the time development of the dynamical variable $h_{ab}(t, \mathbf{x})$ on the three-dimensional hypersurface $\Sigma_t$ along the ``flow of time" given by the vector $t^a=Nn^a+N^a$, where $n^a$ is the unit normal vector field of $\Sigma_t$. $N$ and $N^a$ are not considered dynamical and can be chosen freely since they merely prescribe how to ``move forward in time" (see, e.g., \cite{ Wald:1984rg, Gourgoulhon:2007ue}).

Unlike the homogeneous spacetime described by the FLRW metric \eqref{FLRW}, where the spatial curvature $R^{(3)}$ of $\Sigma_t$ is zero, in the fluctuating spacetime described by the metric \eqref{new metric}, $R^{(3)}$ at each point of $\Sigma_t$ is large and fluctuating. This fluctuation is not arbitrary, and it satisfies the Hamiltonian constraint (see, e.g., \cite{Alan:PDEinGR})
\begin{equation}\label{Hamiltonian constraint}
R^{(3)}-K_{ab}K^{ab}+K^2=2\left(\lambda_B+8\pi G\rho\right),
\end{equation}
where the extrinsic curvature $K_{ab}=\frac{1}{2}\mathcal{L}_nh_{ab}$ is the Lie derivative of $h_{ab}$ with respect to $n^a$, and the mean curvature $K$ is the trace of $K_{ab}$ defined by $K=h^{ab}K_{ab}$.

Since the observed macroscopic spatial curvature of the Universe is very small, the large curvature fluctuation of $R^{(3)}$ at small scales should be able to be averaged out to small value macroscopically. In other words, we need to find a spacetime foliation $\{\Sigma_t\}_{t\in\mathbb{R}}$ for which the macroscopic average $\langle R^{(3)}\rangle\big|_{\Sigma_t}\approx 0$. To do this, we start from an arbitrary initial hypersurface $\Sigma_0$ in our given fluctuating spacetime. Taking spatial average over $\Sigma_0$ on both sides of \eqref{Hamiltonian constraint} gives
\begin{equation}\label{average Hamiltonian constraint}
\langle R^{(3)}\rangle\Big|_{\Sigma_0}=2\lambda_{\mathrm{eff}}+\left\langle K_{ab}K^{ab}-K^2\right\rangle\Big|_{\Sigma_0}.
\end{equation}
So the requirement $\left\langle R^{(3)}\right\rangle\big|_{\Sigma_0}\approx 0$ imposes a constraint on the effective cosmological constant
\begin{equation}
\lambda_{\mathrm{eff}}\approx -\frac{1}{2}\left\langle K_{ab}K^{ab}-K^2\right\rangle\Big|_{\Sigma_0}.
\end{equation}
Expanding the terms $K_{ab}K^{ab}-K^2$ gives
\begin{eqnarray}\label{kab-ksquare}
&&K_{ab}K^{ab}-K^2 \\
=&&\sum_{i\neq j\neq k}M_kK_{ij}^2+\sum_{\{i, j\}\neq\{k, l\}}\left(h^{ik}h^{jl}-h^{ij}h^{kl}\right)K_{ij}K_{kl}, \nonumber
\end{eqnarray}
where
\begin{equation}
M_k=h^{ii}h^{jj}-\left(h^{ij}\right)^2, \quad k\neq i\neq j,
\end{equation}
is the $k$th principal minor of $h^{ab}$. Since by definition the metric matrix $h^{ab}$ is positive definite, we have $M_k>0$.

The extrinsic curvature $K_{ab}$ has six independent components, it describes the rate of change of the spatial metric $h_{ab}$ in the direction orthogonal to $\Sigma_0$. The mean curvature $K$ describes the average property of $K_{ab}$. It is related to the spatial volume element $\sqrt{h}\,dx^1\wedge dx^2\wedge dx^3$ by $K=\frac{1}{\sqrt{h}}\mathcal{L}_n\sqrt{h}$, where $h=\mathrm{det}(h_{ab})$ is the determinant of $h_{ab}$. So, physically, $K>0$ means on average the space is locally expanding, while $K<0$ means on average the space is locally contracting.

Since general relativity is time reversal invariant, for every expanding solution $K>0$ there is a corresponding contracting solution $K<0$. Or more precisely, if $(h_{ab}, K_{ab})$ is an allowed initial data on $\Sigma_0$, so is $(h_{ab}, -K_{ab})$ \cite{Carlip:2018zsk}. Thus, for $\{i, j\}\neq\{k,l\}$, the following four pairs of components
\begin{equation}
(K_{ij}, K_{kl}),\,(K_{ij}, -K_{kl}),\,(-K_{ij}, K_{kl}),\,(-K_{ij}, -K_{kl}) \nonumber
\end{equation}
are equally likely to happen on $\Sigma_0$. Then because, in general, there is no particular relationship between the components of the extrinsic curvature, we have, for the second term in \eqref{kab-ksquare}, the above four cases would statistically cancel each other that the macroscopic spatial average
\begin{equation}\label{zero macroscpic average}
\left\langle\left(h^{ik}h^{jl}-h^{ij}h^{kl}\right)K_{ij}K_{kl}\right\rangle=0, \quad \{i, j\}\neq\{k, l\}
\end{equation}
for a large collection of possible choices of $\Sigma_0$. So only the first term in \eqref{kab-ksquare} survives after the spatial averaging, so that we have
\begin{equation}\label{negative lambdaeff}
\lambda_{\mathrm{eff}}\approx -\sum_{\substack{1\leq i<j\leq 3\\i\neq j\neq k}}\langle M_kK_{ij}^2\rangle\Big|_{\Sigma_0}<0.
\end{equation}

The subsequent evolution of $\langle R^{(3)}\rangle$ depends on the choice of $N$ and $N^a$. For a large negative $\lambda_{\mathrm{eff}}$, once $\Sigma_0$ is chosen to make $\langle R^{(3)}\rangle\big|_{\Sigma_0}\approx 0$, it can be shown that, if we choose the average $\langle N\rangle=1$, $\langle N^a\rangle=0$, the subsequent evolution would keep\footnote{See Sec. IV of \cite{PhysRevD.102.023537} for more details.} $\langle R^{(3)}\rangle\big|_{\Sigma_t}\approx 0$.

The subsequent evolution of the initial data $(h_{ab}, K_{ab})$ is very complicated. Fortunately, the evolution of their average properties described by the determinant $h$ of $h_{ab}$ and the trace $K$ of $K_{ab}$ is relatively simple. Let us generalize the ``global" scale factor $a(t)$ in the FLRW metric \eqref{FLRW} to the ``local" scale factor $a(t, \mathbf{x})$ by letting $h=a^6$. Since the spatial volume element $\sqrt{h}=|a|^3$, $|a(t, \mathbf{x})|$ describes the relative ``size" of space at each point.

Without loss of generality of the metric, we can choose the shift vector $N^a=0$. Then we have $K=\frac{3}{a}\frac{da}{d\tau}$ and $a$ satisfies the evolution equation\footnote{See Sec.III and Sec.V of \cite{PhysRevD.102.023537} for details of the derivation of the evolution equation \eqref{evolution with N}}
\begin{equation}\label{evolution with N}
\frac{d^2a}{d\tau^2}+\left[\frac{1}{3}\left(2\sigma^2-\lambda_{\mathrm{eff}}-\frac{D^aD_aN}{N}\right)+F_{\mathbf{x}}(t)\right]a=0,
\end{equation}
where $d\tau=Ndt$, $\sigma^2>0$ is the shear that characterizes the magnitude of the local anisotropy, $\lambda_{\mathrm{eff}}$ is the effective cosmological constant\footnote{For simplicity, here we have assumed that the vacuum equation of state \eqref{vacuum equation of state} is valid so that the $\lambda_{\mathrm{eff}}$ in \eqref{evolution with N} equals to the $\lambda'_{\mathrm{eff}}$ in Eq.(65) of \cite{PhysRevD.102.023537}. Whether \eqref{vacuum equation of state} is valid or not does not affect our final result.} defined by \eqref{contributions to effective cc new}, and $F_{\mathbf{x}}(t)$ is the vacuum stress energy tensor fluctuation term defined by
\begin{equation}\label{F definition}
F_{\mathbf{x}}(t)=\frac{4\pi G}{3}\left(\rho+h^{ab}T_{ab}-\left\langle\rho+h^{ab}T_{ab}\right\rangle\right).
\end{equation}

Physically, \eqref{evolution with N} describes the dynamical evolution of the local scale factor $a$ measured by the local Eulerian observer who stays at $\mathbf{x}=\mathrm{constant}$. There is a nongravitational force $\mathbf{F}=m\mathbf{D}N/N$ [where $\mathbf{D}=(D_1, D_2, D_3)$] acting on the Eulerian observer to maintain its constant position. The effect of this nongravitational force on the evolution of $a$ is represented by the term $D^aD_aN/N$ in \eqref{evolution with N}. We should use free-falling observers who only feel gravity to measure the local evolution of $a$ so that $a$ represents only the effect of gravity. In other words, we need to choose $N$ to be spatial independent or at least to make the average $\langle D^aD_aN/N\rangle=0$ to make sure the average effects of the nongravitational forces are zero. The simplest choice is $N=1$ and then the metric \eqref{new metric} becomes
\begin{equation}\label{Gaussian normal coordinate}
ds^2=-dt^2+h_{ab}(t, \mathbf{x})dx^adx^b.
\end{equation}
The evolution equation \eqref{evolution with N} reduces to
\begin{equation}\label{full evolution 1}
\ddot{a}+\left(\frac{1}{3}\left(2\sigma^2-\lambda_{\mathrm{eff}}\right)+F_{\mathbf{x}}(t)\right) a=0.
\end{equation}

Next we study dynamics of $a$ given by the evolution equation \eqref{full evolution 1} when $\lambda_{\mathrm{eff}}$ is negative as required by \eqref{negative lambdaeff}. By definition, the term $F_{\mathbf{x}}(t)$, which represents the vacuum stress tensor fluctuations, has zero expectation value. For simplicity, we first exclude this term and study the following equation:
\begin{equation}\label{a evolution no vacuum}
\ddot{a}+\frac{1}{3}\left(2\sigma^2-\lambda_{\mathrm{eff}}\right)a=0.
\end{equation}

\begin{figure}
\includegraphics[scale=0.33]{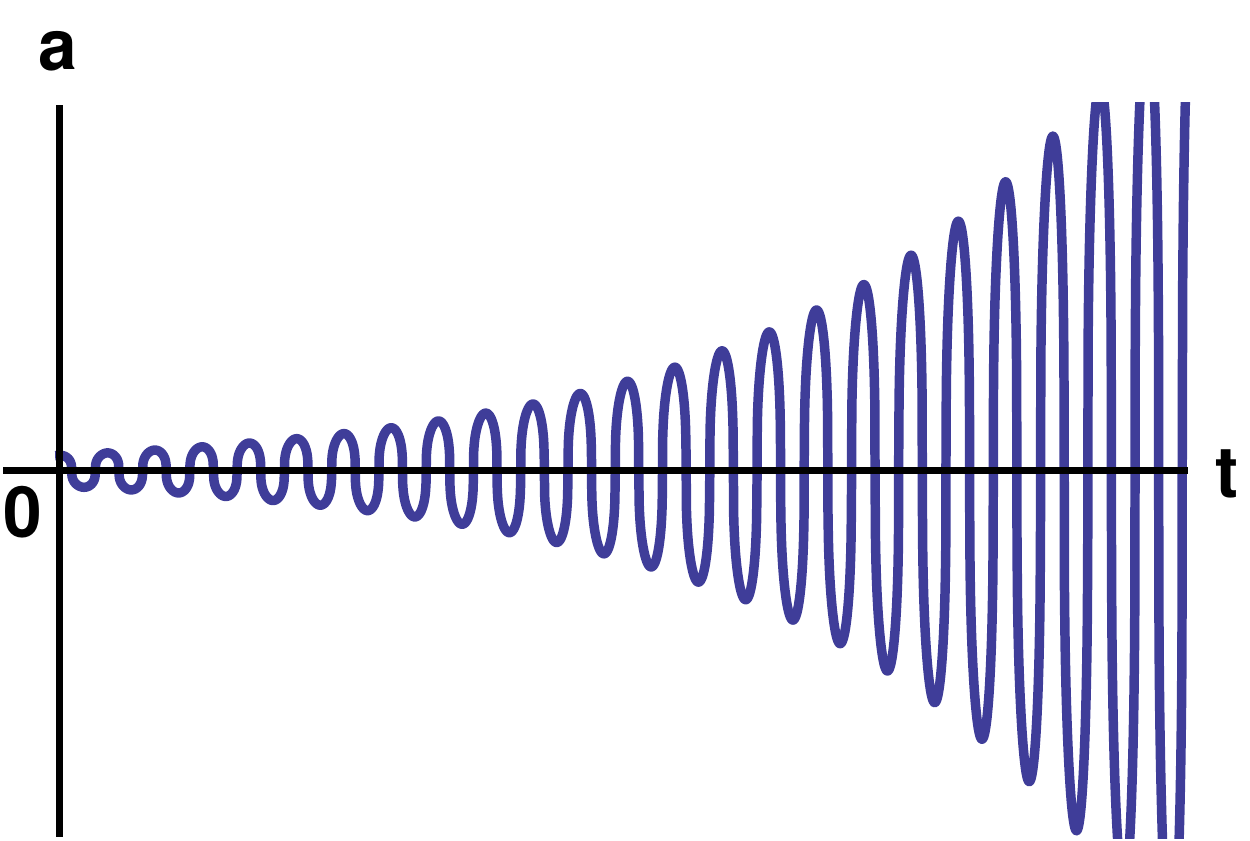}
\includegraphics[scale=0.33]{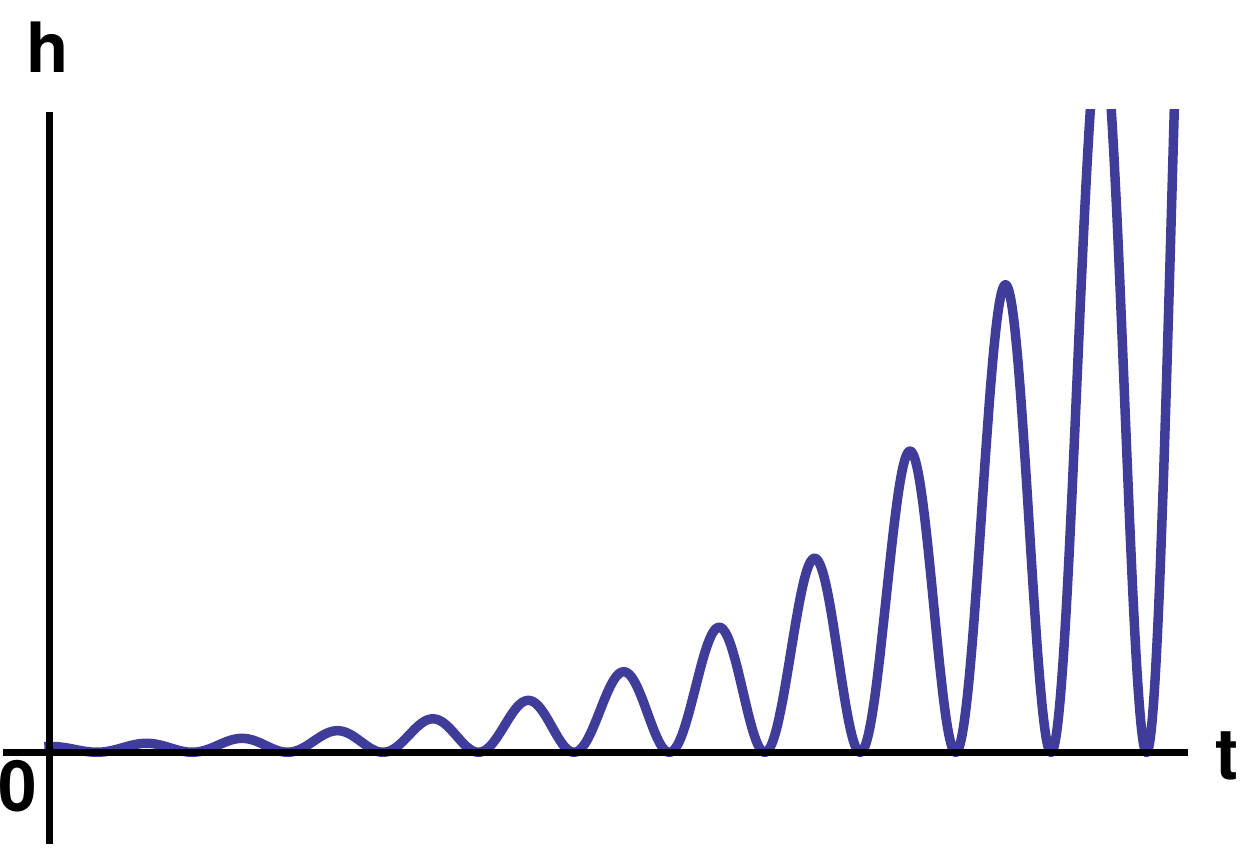}
\caption{\label{schematic plot}Schematic plots of the oscillations of the local scale factor $a$, the local determinant $h=a^6$. As $a$ goes across $0$, $h$ decreases continuously to $0$ and then increases back to positive values. The amplitude of $a$ grows exponentially with a tiny exponent $H=\alpha\Lambda e^{-\beta\frac{\sqrt{-\lambda_B}}{\Lambda}}$ (Eq.\eqref{dependence of H on Lambda}) which gives an accelerating universe.}
\end{figure}

The exact dynamics given by \eqref{a evolution no vacuum} is actually chaotic since the shear $\sigma^2$ satisfies another complicated nonlinear equation [Eq. (72) in \cite{PhysRevD.102.023537}]. However, important qualitative behaviors of $a$ can still be obtained. In fact, since $\sigma^2>0$ and $-\lambda_{\mathrm{eff}}>0$ that $\frac{1}{3}\left(2\sigma^2-\lambda_{\mathrm{eff}}\right)$ must be positive, Eq. \eqref{a evolution no vacuum} describes an oscillator with varying frequency. Thus, the solution for $a$ must be oscillating around zero. In this process, the determinant $h=a^6\geq 0$ decreases continuously to zero and then bounces back to positive values as $a$ crosses zero (see Fig. \ref{schematic plot}). Physically, this means, on average, the space locally collapses to zero size and then immediately bounces back. It will then collapse and expand again and again; i.e., locally, the space is alternatively switching between expansion and contraction\footnote{$a=0$ are actually curvature singularities, it is necessary to assume that the spacetime evolution can continue beyond $a=0$ and the bounces happen. We argue that this continuation is quite natural, at least follows the evolution equation \eqref{a evolution no vacuum} $a$ would not stop at $0$. For more detailed discussions, see Sec. VIII C of \cite{PhysRevD.102.023537}.}.

Since on the initial Cauchy surface $\Sigma_0$, $K>0$ and $K<0$ are equally possible initial data, we have that, in general, the initial conditions $a(0, \mathbf{x})$ and $\dot{a}(0, \mathbf{x})$ for the oscillator equation \eqref{a evolution no vacuum} would take different values at different spatial points. So the phases of these oscillations of $a(t, \mathbf{x})$ at different spatial points would be different. In other words, at any instant of time $t$, the space would be expanding at one point and contracting at neighboring points and vice versa. These phase differences result in a large cancellation between the local expansions and contractions when performing the macroscopic average over the hypersurfaces $\Sigma_t$. Note that the macroscopic average does not require a very large volume: a cubic centimeter contains some $10^{100}$ Planck-size regions. Therefore, we have the average $\langle K\rangle$ over $\Sigma_t$ approaches zero for any sensible macroscopic average procedure. The observed macroscopic volume of the space would then approach a constant
\begin{equation}
V(t)=\int d^3x\sqrt{h}=\int d^3x|a|^3=\mathrm{constant}.
\end{equation}
Thus in this spacetime the large cosmological constant $\lambda_{\mathrm{eff}}$ has a huge effect at small scale but becomes hidden macroscopically.

Now we study the effect of the vacuum stress tensor fluctuation by putting the $F_{\mathbf{x}}(t)$ term back. Its magnitude goes as $F_{\mathbf{x}}(t)\sim G\Lambda^4$. If the term $-\lambda_{\mathrm{eff}}=-\lambda_B-8\pi G\langle\rho\rangle$ in \eqref{full evolution 1} is dominant over $F_{\mathbf{x}}(t)$, i.e., if $-\lambda_B\gg\Lambda^2\geq G\Lambda^4$ (assuming $\Lambda\leq E_P$), the term $F_{\mathbf{x}}(t)$ would just serve as a perturbation on the dynamics of $a(t, \mathbf{x})$ determined by \eqref{a evolution no vacuum}. This perturbation should be unstable. A weak parametric resonance effect is expected to happen that the perturbed solution to \eqref{full evolution 1} is asymptotic to\footnote{For more technical details, see Sec.VII of \cite{PhysRevD.102.023537}.}
\begin{equation}
a(t, \mathbf{x})\sim e^{Ht}a_0(t, \mathbf{x}),\quad H>0,
\end{equation}
where $a_0(t, \mathbf{x})$ is the solution to \eqref{a evolution no vacuum}. Then the observed macroscopic volume of the space would be
\begin{equation}
V(t)=\int d^3x|a|^3=e^{3Ht}V(0).
\end{equation}
So $H$ represents the Hubble expansion rate produced by the vacuum stress energy tensor fluctuations. This produces a slowly accelerated expanding universe. The magnitude of $H$ is related to the bare cosmological constant $\lambda_B$ and the high energy cutoff $\Lambda$ by\footnote{See Sec.VII of \cite{PhysRevD.102.023537} for the derivation of the result \eqref{dependence of H on Lambda}.}
\begin{equation}\label{dependence of H on Lambda}
H=\alpha\Lambda e^{-\beta\frac{\sqrt{-\lambda_B}}{\Lambda}},
\end{equation}
where $\alpha, \beta>0$ are two dimensionless constants whose values depend on the detailed fluctuation property of $F_{\mathbf{x}}(t)$.

In this fluctuating spacetime, the $\lambda_{\mathrm{eff}}$ defined by \eqref{contributions to effective cc new} is large and negative; it is not a macroscopically observable quantity. From \eqref{dependence of H on Lambda} we obtain a new macroscopically observable effective cosmological constant
\begin{equation}\label{lambda dependence on bare lambda}
\lambda^{(\mathrm{obs})}_{\mathrm{eff}}=3H^2=3\alpha^2\Lambda^2 e^{-2\beta\frac{\sqrt{-\lambda_B}}{\Lambda}}.
\end{equation}
Because of the exponential suppression, $\lambda^{(\mathrm{obs})}_{\mathrm{eff}}$ is naturally small if $\sqrt{-\lambda_B}/\Lambda$ is large, so no fine-tuning of $\lambda_B$ to the accuracy of $10^{-122}$ is needed.

\section{Conclusion}
Based on the uncertainty principle of quantum mechanics and the equivalence principle of general relativity, both the quantum vacuum and the spacetime metric at small scales (Planck scale) are highly inhomogeneous and anisotropic. The standard formulation of the cosmological constant problem is based on the homogeneity and isotropy assumption. So, logically speaking, once this assumption on which the cosmological constant problem is based is removed, the problem itself is gone.

One needs to reformulate this problem by replacing the homogeneity assumption with the inhomogeneous one. The work presented in this Letter is such a reformulation. We have shown that once the inhomogeneity at small scales is properly considered, the cosmological constant problem does not arise, since the large gravitational effect of the quantum vacuum is hidden at small scale spacetime fluctuations. The fine-tuning of the cosmological constant is no longer needed. The stress energy tensor fluctuations of the quantum fields vacuum could serve as dark energy to drive the accelerating expansion of the Universe through a weak parametric resonance effect. The physical picture is a more detailed version of Wheeler's spacetime foam \cite{PhysRev.97.511, Wheeler:1957mu, misner1973gravitation} -- at extremely small scales (Planck scale), spacetime fluctuates like foamy and jittery bubbles for which we name \emph{microcyclic} universes that exhibit eternal oscillations of expansions and contractions \cite{PhysRevD.102.023537}.

One issue of this reformulation is the singularities at $a=0$. The occurrence of singularities is an ubiquitous problem of classical general relativity, and not a particular problem of this reformulation. Still under the classical general relativity framework, we have provided a natural prescription to continue the spacetime evolution across the singularities (see Sec. VIII C of \cite{PhysRevD.102.023537} for more detailed discussions). However, a satisfactory quantum theory of gravity is still needed to better understand the nature of the singularities.

\section*{ACKNOWLEDGMENTS}
I would like to thank William G. Unruh for his support of this research and valuable discussions. I would also like to thank Andrei Barvinsky, Mark Van Raamsdonk, Joanna Karczmarek, Philip C. E. Stamp, Samuel Cree, Yin-Zhe Ma, Jian-Wei Mei, Zong-Kuan Guo and Rong-Gen Cai for helpful discussions. This research was funded by the Natural Sciences and Engineering Research Council of Canada through William G. Unruh.

\bibliographystyle{unsrt}
\bibliography{how_vacuum_gravitates}

\end{document}